\documentclass[graphicx,twocolumn]{revtex4-1}
\usepackage{epsfig}
\usepackage{graphics}\graphicspath{{img/}{}}
\usepackage{amssymb,amsmath,amsfonts}
\usepackage{time}
\usepackage{epstopdf}
\usepackage{color}
\usepackage{hyperref}
\hypersetup{
  colorlinks,
  citecolor=magenta,
  linkcolor=blue}

\renewcommand{\vec}[1]{\textnormal{\boldmath$#1$}}

\usepackage[USenglish]{babel}

\newcommand{\be}{\begin{equation}}
\newcommand{\ee}{\end{equation}}

\newcommand{\bc}{\begin{center}}
\newcommand{\ec}{\end{center}}
\newcommand{\bea}{\begin{eqnarray}}
\newcommand{\eea}{\end{eqnarray}}

\renewcommand{\vec}[1]{\textnormal{\boldmath$#1$}}

\newcommand{\ret}{{\rm ret}}

\usepackage{xstring}

\begin{document}

\title{Centripetal Transverse Wakefield in Relativistic Beam}

\author{Gennady Stupakov} 
\address{SLAC National Accelerator Laboratory, Menlo Park, CA, USA}

\begin{abstract}

The effect of the transverse self-force in a relativistic beam has been studied in the earlier paper of the author~\cite{stupakov99_1}. However, the analysis of~\cite{stupakov99_1} missed an important observation and has lead to an incorrect estimate of the emittance growth of the beam when it passes through a long bending magnet. Here we correct that analysis. In particular, we conclude that the emittance growth due to the transverse self-force in a long bend is always much smaller than the emittance growth due to the longitudinal coherent synchrotron radiation (CSR) wake.

\end{abstract}
\maketitle

\section{Introduction}

Many of the basic features of the coherent synchrotron radiation (CSR) of short bunches and its effect on beam dynamics in accelerators are now well established \cite{iogansen60r,murphy95kg,derbenev95rss,murphy97kg}. The effect is usually described in terms of the longitudinal force, or wakefield, that causes the energy loss in the beam, and also redistributes the energy between the particles by accelerating the head and decelerating the tail of the bunch. Coherent radiation becomes most important for short bunches and high currents. More subtle features of CSR such as transition effect due to the entrance to and exit from the bend \cite{saldin97sy}, CSR force in the undulator \cite{saldin98sy}, and shielding due to the close metallic boundaries \cite{murphy97kg,warnock90m,ng90} have also been  studied.

In addition to the longitudinal, there is a transverse force in a short bunch moving on a circular orbit. The problem has been treated in several papers beginning from R. Talman's  work \cite{talman86}, who pointed out that the centrifugal force of  a rotating bunch can result in a noticeable tune shift of betatron oscillations. Later, an important correction to the Talman paper has been made in Ref. \cite{lee88}, where it was shown that due to the energy variation in the bunch, the effect of the transverse force proportional to  $R^{-1}$ is cancelled, and the residual effect is of the order of  $R^{-2}$, that is much smaller than originally predicted. In Ref.~\cite{derbenev96s}, Derbenev and Shiltsev  found  a centripetal force of the order of $R^{-1}$ that differs from Talman's result by a logarithmic factor.

Further studies of the curvature induces transverse force by R. Li and Ya. Derbenev~\cite{rui_li02,li_derb02,Li:2005xy,rui_li08} and by G. Geloni et al.~\cite{Geloni:2002zj,Geloni:2003qt,geloni04etal} has lead to somewhat conflicting conclusions regarding the cancellation effect. This controversy has been emphasized in Ref.~\cite{Geloni:2003qt}.

In Ref.~\cite{stupakov99_1} this author tried to elucidate the cancellation effect applying it first to a coasting beam moving in a constant magnetic field. In this model, the cancellation can be easily established through direct calculation of the beam fields and potentials. The method developed for a coasting beam was then generalized for a bunch moving in a circular orbit. Unfortunately, in the calculation of the deflection angle caused by the transverse force, the cancellation effect was neglected, which resulted in an incorrect estimate of the emittance growth of a bunch passing through a bending magnet. In this paper, we correct the error made in Ref.~\cite{stupakov99_1}. For the sake of completeness of the presentation, below we reproduce a substantial part of the text from Ref.~\cite{stupakov99_1}.

Throughout this paper we assume  ultra-relativistic beam, $v=c$, moving on a circular orbit of radius $R$.

\section{Lienard-Wiechert Potentials and Fields}

\begin{figure}[ht]
    \begin{center}
    \includegraphics[scale=0.6]{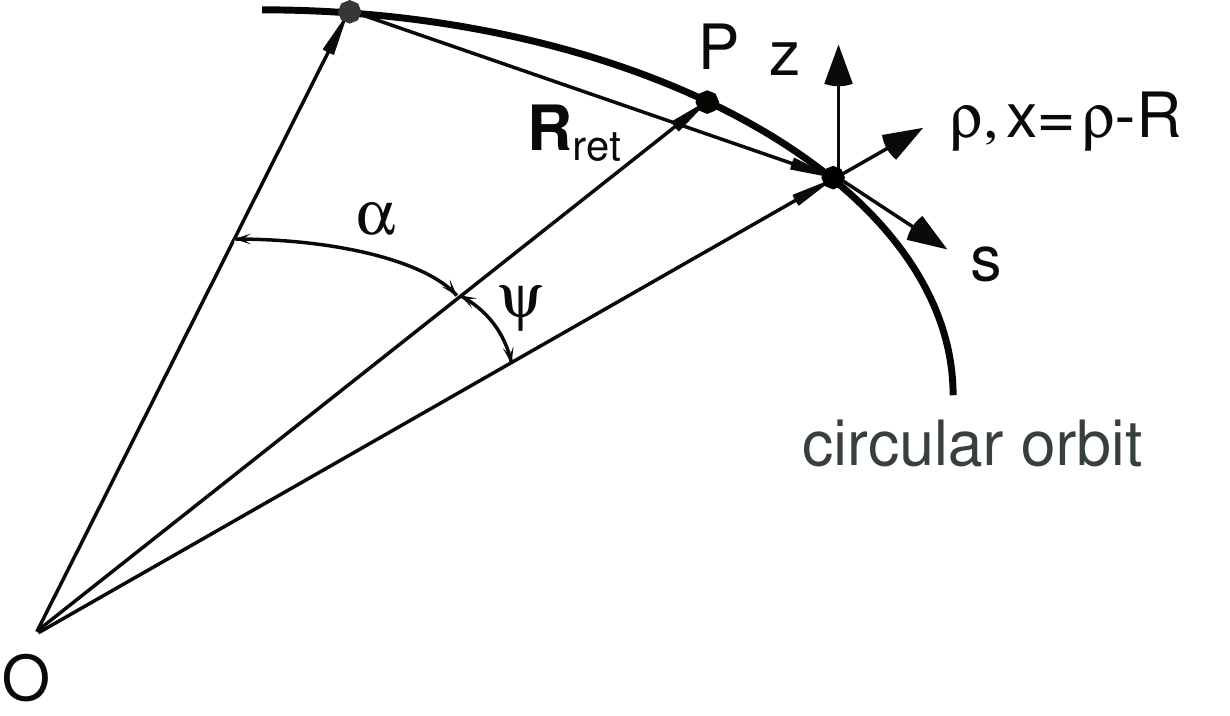}
    \caption{The cylindrical coordinate system $\rho$, $\psi$, $z$ and a circular orbit of radius $R$. The longitudinal coordinate along the orbit is measured by coordinate $s$. Shown are locations of the particle at time
    $t$, the point P,
    the observation point in front of the particle at an angle
    $\psi$,
    and
    the point where the radiation occurred at the retarded time
    (at an angle $\alpha$ behind the particle).
    Vector $\vec{R}_{\ret}$ connects the radiation point with the observation point.}
    \label{coord_system}
    \end{center}
    \end{figure}

The electromagnetic field of  a point charge moving in vacuum, as is well known, can be found using Lienard-Wiechert potentials, and the fields can be explicitly expressed in terms of particle's velocity and acceleration at the retarded time \cite{jackson}. We will use the coordinate system shown in Fig. \ref{coord_system}.
In the limit $v=c$, the fields of the charge are given by the following equations,
    \begin{align} \label{elec_field}
    \vec{E}
    &=
    {q \over c}\frac{\vec{R}_{\ret}\times\left[
    (\vec{R}_{\ret}-R_{\ret}\vec{\beta}_{\ret})\times
    \dot{\vec{\beta}}_{\ret}\right]}
    {\left({R}_{\ret}-\vec R_{\ret}\cdot\vec{\beta}_{\ret}\right)^3},
    \nonumber\\
    \vec{H}
    &=
    {1 \over R_{\ret}}[\vec{E}\times \vec{R}_{\ret}],
    \end{align}
where the  distance $\vec{R}_{\ret}$ shown in Fig. \ref{coord_system} connects the position of the particle at the radiation time $t_\ret$ and the observation point, $R_{\ret} = c(t-t_\ret)$, $\vec{\beta}_{\ret}$ is the unit vector ($v=c$) directed along the particle velocity at the radiation time, and $\dot{\vec{\beta}}_{\ret}$ is the derivative of the velocity at that time. To these fields correspond the following scalar and vector potentials,
    \begin{align}
    \phi
    =
    \frac{q}
    {\left({R}_{\ret}-\vec R_{\ret}\cdot\vec{\beta}_{\ret}\right)}
    ,\qquad
    \vec A
    =
    \frac{q\vec{\beta}_{\ret}}
    {\left({R}_{\ret}-\vec R_{\ret}\cdot\vec{\beta}_{\ret}\right)}
    .
    \end{align}

\begin{figure}[!ht]
    \bc
    \includegraphics[scale=0.35]{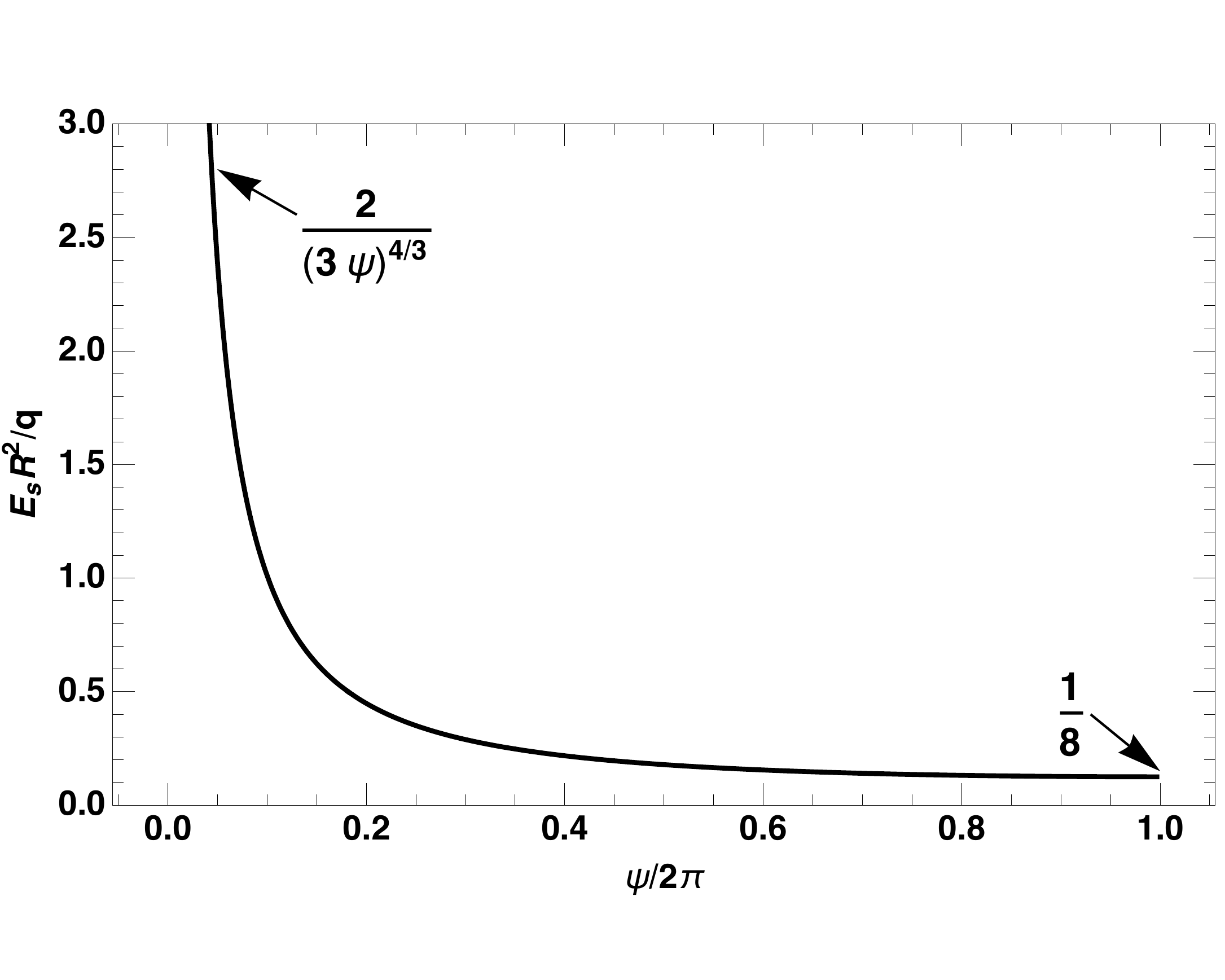}
    \caption{Longitudinal electric field on a circular orbit as
    a function of angle $\psi$ counted from the
    location of the particle.
    The arrows show asymptotic expressions for the field in
    the vicinity of the particle.}\label{long_field}
    \ec
    \end{figure}

If the transverse size of the bunch $\sigma_r$ is much smaller than its length $\sigma_z$, only the field on the orbit interacts with the beam. In this case, the observation point can be chosen on the circle, as shown in Fig.~1, and Eqs.~(\ref{elec_field}) can be written as (see Fig.~\ref{coord_system} for the notations)
    \begin{align} \label{Es}
    E_s(\psi)
    &=
    {q\over R^2}{2 \sin{{1\over 2}(\alpha+\psi)} \over
    [\alpha-\sin(\alpha+\psi)]^3}
    \nonumber \\
    &\times
    \left\{
    2 (\sin{{1\over 2}(\alpha+\psi)})^3
    \left[\cos{{1\over 2}(\alpha+\psi)}-\cos(\alpha+\psi)\right]
    \right.
    \nonumber
    \\
    &-
    \left.\sin(\alpha+\psi)[\alpha-\sin(\alpha+\psi)]
    \right\},
    \nonumber
    \\
    E_\rho(\psi)
    &={q\over R^2}{2 \sin{{1\over 2}(\alpha+\psi)} \over
    [\alpha-\sin(\alpha+\psi)]^3}
    \nonumber \\
    \times&
    \left\{
    2 (\sin{{1\over 2}(\alpha+\psi)})^3
    \left[\sin{{1\over 2}(\alpha+\psi)}-\sin(\alpha+\psi)\right]
    \right.
    \nonumber
    \\
    +&
    \left.\cos(\alpha+\psi)[\alpha-\sin(\alpha+\psi)]
    \right\},
    \nonumber
    \\
     H_z(\psi)
     &=
     E_s \sin{{1\over 2}(\alpha+\psi)}
    -E_\rho \cos{{1\over 2}(\alpha+\psi)},
    \end{align}
where $E_s(\psi)$ is the longitudinal and $E_\rho(\psi)$ -- radial
components of the electric field, $H_z(\psi)$ is the vertical
magnetic field, $\psi = s/R$, and the angle $\alpha$ is related to
the position of the observation point by equation
    \be \label{angle_alpha}
    \alpha = 2|\sin{1\over 2}(\psi+\alpha)|.
    \ee
The plot of the longitudinal field as a function of the position on
the circle is shown in Fig. \ref{long_field}.

The potential $\phi$ on the orbit is given by the following equation
    \begin{align}\label{eq:5}
        \phi
        =
        \frac{q}{R}
        \frac{1}{\alpha-\sin(\alpha+\psi)}
        .
    \end{align}

The electric field $E_s$ per unit charge is equal to the longitudinal
wake $w$. For small distances, $s \ll R$, one finds from Eqs.
(\ref{Es}) and (\ref{angle_alpha}) ,
    \be
    w(s)={1\over q}E_s(s)\approx \left\{
    \begin{array}{cl}
    {2(3s)^{-4/3} R^{-2/3}}, & s>0, \\
    0, &                            s<0.
    \end{array}
    \right.
    \ee
For a short bunch ($\sigma_z \ll R$) with a given charge distribution
$\lambda(s)$ ($\int \lambda(s)ds = 1$), the longitudinal wake for the
bunch  is defined as a convolution with the distribution function,
    \begin{align*}
    &w_\mathrm{bunch}(s)
    =
    \int_{-\infty}^{\infty}w(s-s')\lambda(s')ds'
    \\
    &=
    {2\over 3^{4/3}R^{2/3}}\int_s^\infty{\lambda(s') ds' \over
    (s'-s)^{4/3}}.
    \end{align*}
The problem here is that the above integral diverges when $s' \rightarrow s$. This problem is overcome if one integrates the last equation by parts (neglecting the non-integral term!):
    \be \label{bunch_wake}
    w_{\rm bunch}(s)=
    {2\over 3^{1/3}R^{2/3}}\int_s^\infty{ ds' \over
    (s'-s)^{1/3}} {d\lambda(s') \over ds'}.
    \ee
Now the integral converges, and gives the right result for the wake \cite {murphy95kg,derbenev95rss}. The justification for this integration by parts can be found in a more accurate treatment of the fields in a small vicinity of the particle \cite{murphy97kg,saldin97sy}.

We also mention here that although we obtained the above result
assuming an infinitely thin beam, the applicability condition for the
longitudinal wake is actually very mild, ${\sigma_r / \sigma_z} \ll
\left({R /\sigma_z}\right)^{1/3}$ \cite {derbenev95rss}.
    \begin{figure}[!ht]
    \bc
    \includegraphics[scale=0.35]{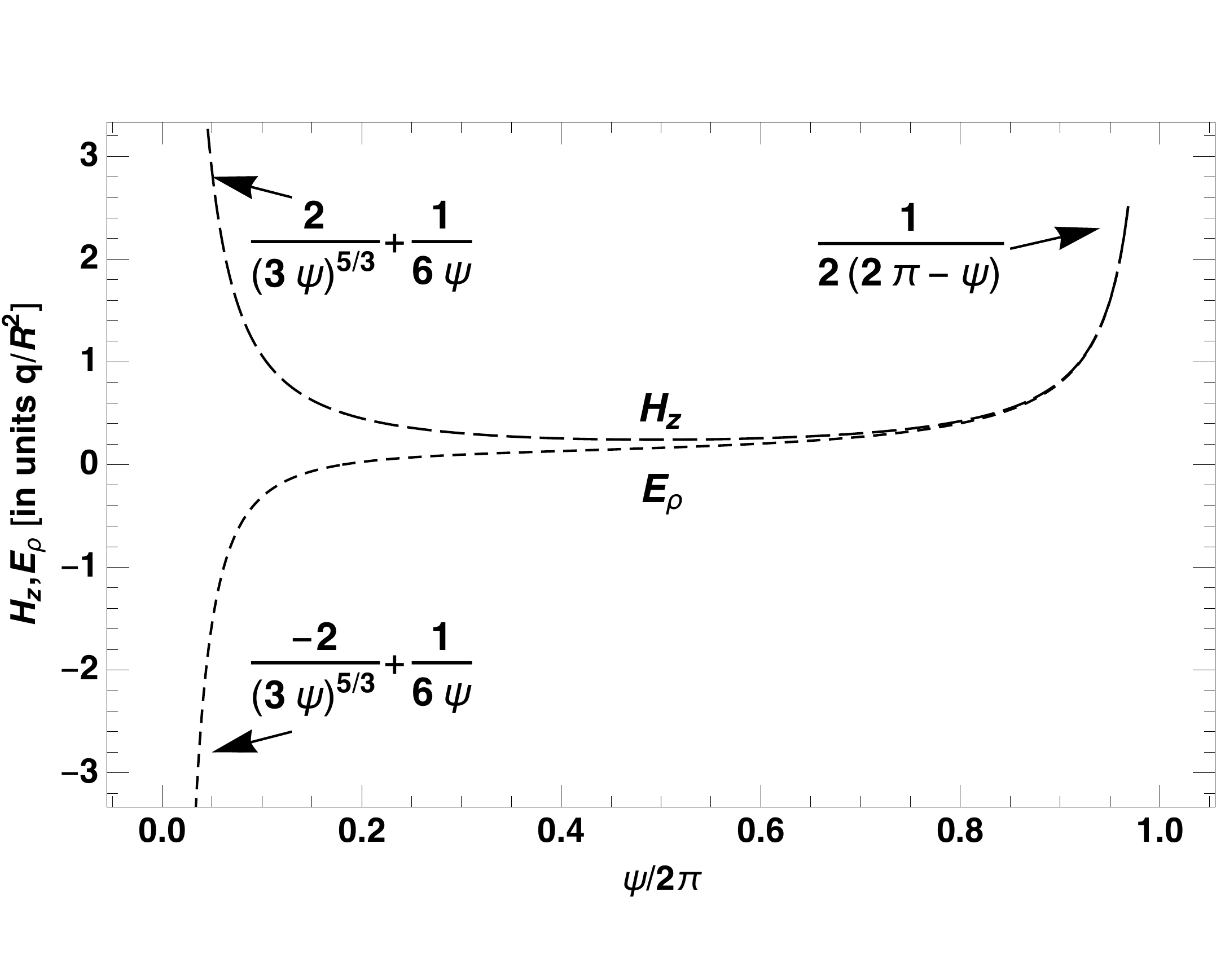}
    \ec
    \caption{Radial electric field $E_\rho$ and
    vertical magnetic field $H_z$ on the orbit.
    Arrows show the asymptotic expressions for the fields near
    the particle.} \label{rad_elec_field}
    \end{figure}
    \begin{figure}[h!t]
    \bc
    \includegraphics[scale=0.35]{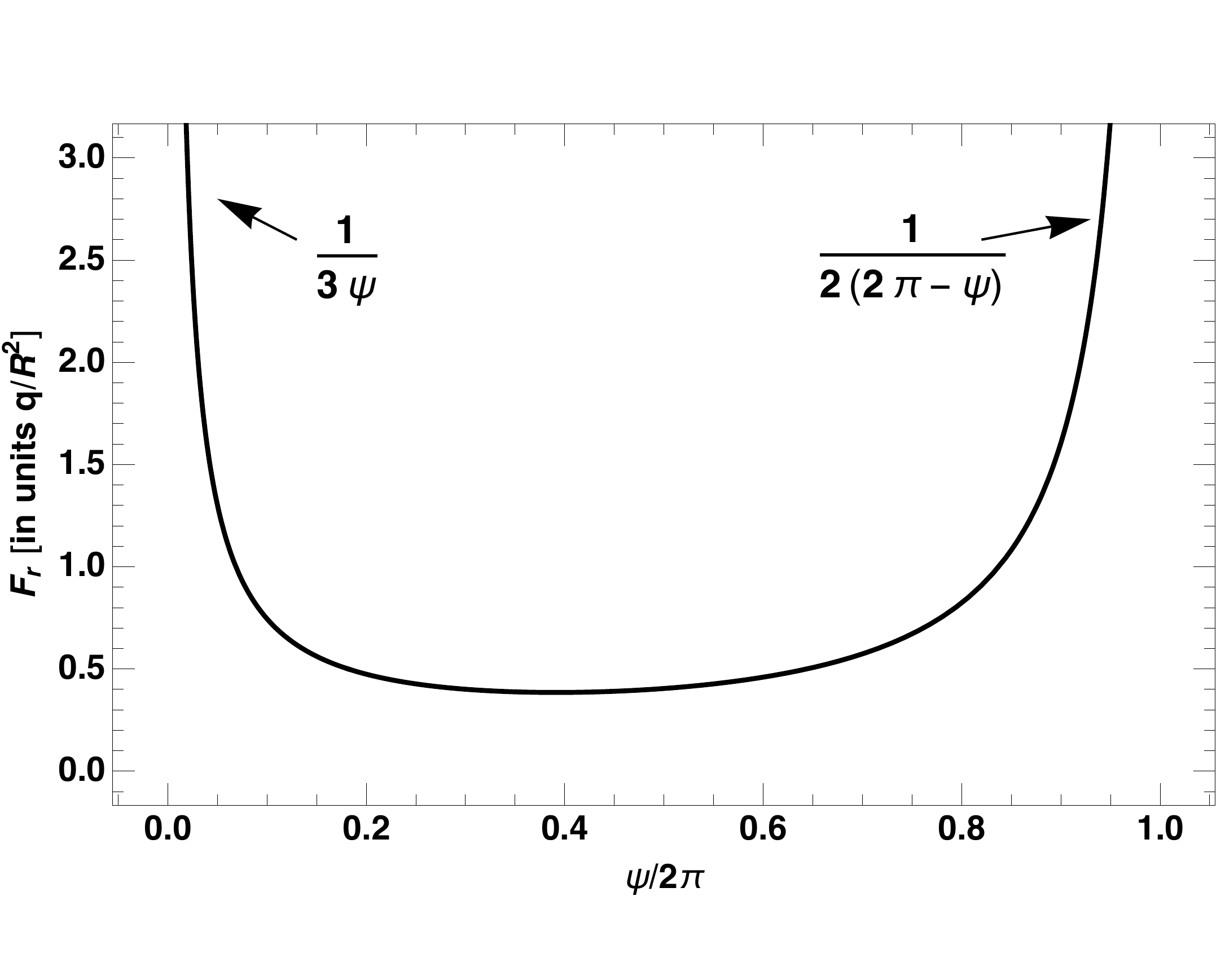}
    \caption{Transverse force per unit charge $F_\rho$ as a function of
    angle $\psi$. Arrows show  asymptotic expressions for the force
    near the particle.} \label{trans_force}
    \ec
    \end{figure}

At this point we are tempted to apply the same approach for the calculation of the transverse force in a thin bunch. The transverse force (per unit charge) $F_\rho$ for an ultrarelativistic bunch is
$F_\rho=E_\rho+H_z$. The plots of $E_\rho$ and $H_z$ are shown in Fig. \ref{rad_elec_field}, and the transverse force as a function of angle $\psi$ on the circular orbit is shown in Fig.~\ref{trans_force}. Asymptotically, for small positive $\psi$ in front of the particle $F_\rho \approx q/3R^2\psi$; behind the particle, for negative small $\psi$, $F_\rho \approx q/R^2|\psi|$. Again, if we want to convolute this force with the bunch distribution and to find the transverse wakefield for the bunch, as we did above for the longitudinal wake, the integral would logarithmically diverge. As we will see in the next section, there is a profound reason for such divergence: the transverse force depends on  the beam radius $\sigma_r$ that we neglected in the above consideration. We will also see that the force $F_\rho$ enters into the equations of the beam transverse dynamics in combination $F_\rho-\phi/R$. A remarkable fact that was missed in Ref.~\cite{stupakov99_1} is that if one calculates the potential $\phi$ using Eq.~\eqref{eq:5}
one finds that the difference $F_\rho-\phi/R$ is exactly equal to zero,
    \begin{align}\label{eq:8}
        F_\rho
        -
        \frac{\phi}{R}
        =
        0.
    \end{align}
Strictly speaking, this relation holds everywhere on the circular orbit except for the point where the point charge is located --- at this point, both terms in this equation are infinite, and no conclusion can be made in regard of their difference. As we will see below, after analysis of the fields in a beam of a finite transverse size, a more accurate expression for this difference contains a delta function localized at the location of the point charge.

\section{ Transverse Force -- Coasting Beam }

\begin{figure}[h]
    \bc
    \includegraphics[scale=0.5]{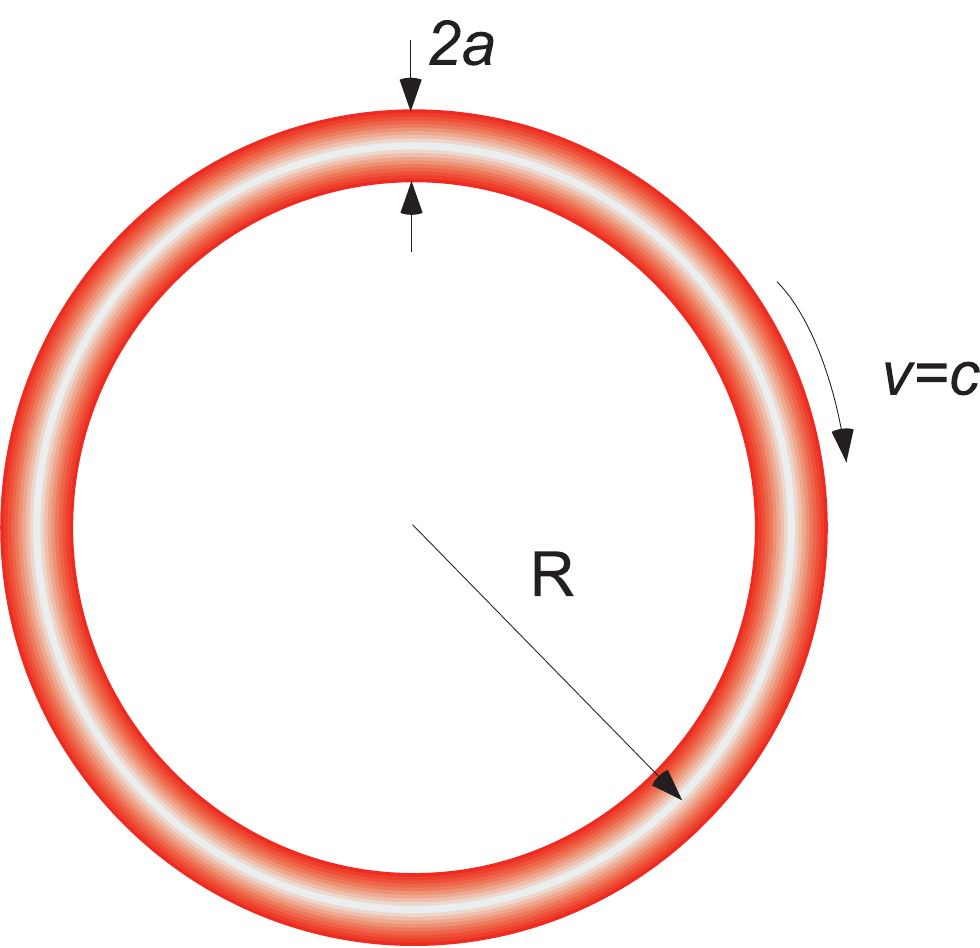}
    \caption{Coasting beam of radius $a$ moving along a circular orbit
    in vacuum.} \label{dc_beam}
    \ec
    \end{figure}

To make our consideration of the transverse force as simple as
possible we begin here from a problem of a coasting relativistic dc
beam of radius $a$,  shown in Fig. \ref{dc_beam}. To find the
transverse force $f_\rho$  acting on a unit length of the beam  in
this case, we will use the energy principle that relates the force to
the variation of the energy of the system under infinitesimally small
displacement \cite{landau_lifshitz_ecm}.
Since the electric and magnetic fields of a coasting beam are time
independent, the electromagnetic energy of the beam is the sum of the
electrostatic and magnetic energies. To find them, we need to know
the capacitance and inductance of a charged rotating ring, which can
be found in textbooks on electrodynamics (see, e.g.,
\cite{landau_lifshitz_ecm}).

The inductance of a circular current in vacuum is given by the
following formula,
    \be \label{inductance}
    L=4\pi R\left(\ln{8R \over a}-{7\over 4}\right).
    \ee
Differentiating the magnetic energy $LI^2/2c^2$ with respect to the
circumference of the beam, we obtain the radial magnetic force per
unit length of the ring,
    \be \label{magn_force}
    f_m={\partial (LI^2/2c^2) \over \partial (2\pi R)}=
    {I^2 \over Rc^2 }\left(\ln{8R \over a}-{3\over 4}\right).
    \ee

Analogously, the capacitance $C$ of the uniformly charged ring is
    \be \label{capacitance}
    C^{-1}={1 \over \pi R}\left(\ln{8R \over a}+{1\over 4}\right),
    \ee
and the electric force per unit length is equal to the derivative of
the electrostatic energy with respect to the circumference (with the
minus sign),
    \be \label{elect_force}
    f_e=-{\partial (Q^2/2C) \over \partial (2\pi R)}=
    {Q^2 \over 4\pi^2 R^3 }\left(\ln{8R \over a}-{3\over 4}\right).
    \ee
Taking into account that for an ultrarelativistic beam $Q=2\pi R I/
c$, we find that the electric and magnetic forces are equal, which is
expected in the limit $v=c$.

Adding Eq. (\ref{magn_force}) and (\ref{elect_force}) gives the total
force,
    \be \label{total_force_ring}
    f_\rho=f_m+f_e=
    {I^2 \over Rc^2 }\left(2\ln{8R \over a}-{3\over 2}\right).
    \ee

We can also easily find the force, if the beam propagates in  a
circular pipe of radius $b$. In this case the capacitance of the ring
is,
    \be
    C^{-1}={1 \over \pi R}\left(\ln{b \over a}+{1\over 4}\right),
    \ee
with the electric force
    \be
    f_e=-{\partial (Q^2/2C) \over \partial (2\pi R)}=
    {Q^2 \over 4\pi^2 R^3 }\left(\ln{b \over a}-{3\over 4}\right).
    \ee
Calculation of the magnetic force in this case shows that, as above,
it is equal to the electric one, and the total force is twice the
electric force,
    \be \label{force_in_pipe}
    f_\rho=f_m+f_e=2f_e=
    {I^2 \over Rc^2 }\left(2\ln{b \over a}-{3\over 2}\right).
    \ee

The above derivation is very simple, but it does not tell how the
centrifugal force varies in the cross section of the beam. To answer
this question, we need to find the electric and magnetic fields
inside the beam. For the beam in vacuum, we will find these fields
using a perturbation theory in small parameter $a/R$.

As a first step in calculations, we need the electrostatic potential
$\phi$ and the vector potential $A_\theta$ at  distances far from the
center of the beam in comparison with the beam radius,
$r=\sqrt{x^2+z^2} \gg a$ but close, relative to the orbit size, $R
\gg r$. This approximation is equivalent to the limit of infinitely
thin beam, $a \rightarrow 0$, and the result can be found in
textbooks (see, e.g. \cite{landau_lifshitz_ecm}),
    \be
    \phi(\rho,z)=2\mu \sqrt{R \over \rho}\kappa K(\kappa^2),
    \ee
    \be
    A_\theta(\rho,z)=\left({2\over \kappa^2}-1\right)
    \phi-4\mu \sqrt{R \over \rho}\kappa^{-1} E(\kappa^2),
    \ee
where $ \kappa^2=4 \rho R / [(\rho+R)^2+z^2]$, $\rho$ is the radius
counted from the center of the orbit, and $\mu$ is the charge per
unit length. Expanding this expression in the vicinity of the beam,
$x=\rho-R$, $r=\sqrt{x^2+z^2} \ll R$, one finds
    \be \label{phi_asymp}
    \phi=-2\mu \left(1-{x\over 2R}\right)\ln{\sqrt{x^2+z^2} \over
    8R}+\mu{x \over R}+\ldots,
    \ee
    \be \label{A_asymp}
    A_\theta=\phi -4\mu\left(1-{x\over 2R}\right) +\ldots.
    \ee

To find the  potential $\phi$ inside the beam we need to solve the
Poisson equation,
    \be \label{eq_for_phi}
    \Delta \phi ={1 \over \rho}{\partial \over \partial \rho} \rho {\partial \phi \over \partial
    \rho}+{\partial^2 \phi \over \partial
    z^2}=-4\pi e n(\rho,z),
    \ee
where $n$ is the particle density in the beam. Using the coordinate
$x$, $ \rho = R+x, \,\, x \ll R $, we can expand the first term in
the equation keeping only linear terms in $R^{-1}$,
    \be
    {\partial^2 \phi \over \partial x^2}+{\partial^2 \phi
    \over \partial z^2}=
    -4\pi e n-{1 \over R}{\partial \phi \over \partial x}+O(R^{-2}).
    \ee
Let us assume that $\phi = \phi_0+\phi_1$, where $\phi_0$ is the
potential in the limit $R \rightarrow \infty$, and $\phi_1$ is linear
in $R^{-1}$, $\phi_1 \ll \phi_0$. In the zeroth approximation, we
have
    \be \label{poisson_eq_zeroth}
    {\partial^2 \phi_0 \over \partial x^2}
    +{\partial^2 \phi_0 \over \partial
    z^2}=-4\pi e n.
    \ee
For a constant density beam, $n = {\rm const}$ for $r<a$, the
solution of Eq. (\ref{poisson_eq_zeroth}) is
    \bea \label{phi0}
    \phi_0&=&\mu\left(1-{r^2\over a^2}\right)+2\mu \ln \frac{8R}{a}, \qquad r <a, \nonumber \\
    \phi_0&=&-2\mu\ln{r\over 8R}, \qquad    r>a,
    \eea
where $\mu = \pi a^2en$ is the beam charge per unit length. In Eq.
(\ref{phi0}) we took into account that in the region $a \ll r \ll
R$ it should match the axisymmetric part of the asymptotic
expression $\phi \approx 2\mu \ln (8R/r)$ which follows from Eq.
(\ref{phi_asymp}). In the first order, the potential $\phi_1$
satisfies the equation
    \be \label{eq_for_phi_1}
    {\partial^2 \phi_1 \over \partial x^2}+{\partial^2 \phi_1 \over \partial
    z^2}=-{1 \over R}{\partial \phi_0 \over \partial x}.
    \ee
The solution can be found by solving Eq. (\ref{eq_for_phi_1}) and
using as a boundary condition the  asymptotic behaviour for large
$r$, given by Eq. (\ref{phi_asymp}). The potential $\phi_1$ inside
the beam is,
    \be \label{phi_in_beam}
    \phi_1={\mu\over R}
    x\left(
    1+{r^2\over 4a^2}
    +\ln{a\over 8R}
    \right), \qquad r <a.
    \ee

In a similar fashion, we can find the vector potential $A_\theta$
that satisfies the equation
    \be \label{eq_for_A}
    \Delta A_\theta = -4\pi e n,
    \ee
but has a different asymptotic condition at large $r$, Eq.
(\ref{A_asymp}). The result is
    \bea \label{a_theta_in_beam}
    A_{\theta 0}&=&-\mu\left({r^2\over a^2}+2 \ln{a\over 8R}-3\right), \qquad r
    <a,\nonumber \\
    A_{\theta 1}&=&{\mu\over R}
    x\left(
    3+{r^2\over 4a^2}
    +\ln{a\over 8R}
    \right), \qquad r <a.
    \eea

It is interesting to note, that although $\phi$ and $A_{\theta}$
satisfy the same equation (see Eq. (\ref{eq_for_phi}), and Eq.
(\ref{eq_for_A})), $A_\theta \ne \phi$ due to different asymptotic
conditions at large $r$.

Using Eqs. (\ref{phi_in_beam}) and Eq. (\ref{a_theta_in_beam}) for
the potentials, one can find the fields inside the beam and calculate
the distribution of the transverse force over the cross section of
the bunch. This force, per unit charge, as a function of radius $r$ is given by the
following equation,
    \be \label{F_rho_inside}
    F_\rho=
    {\mu \over R }\left(-1-{r^2\over a^2}+2\ln{8R \over
    a}\right).
    \ee
We see that the force has a parabolic profile with the maximum value
on the axis of the beam.  Averaging this  force over the cross
section yields
    \be \label{force_in_circular_beam}
    \bar{ F}_\rho  =
    {\mu \over R }\left(2\ln{8R \over a}-{3\over 2}\right).
    \ee
To compare this result with Eq. (\ref{total_force_ring}), we need to
take into account that $\mu=I/c$ and the force per units length of
the beam $f_\rho$ equals  the force per unit charge $F_\rho$
multiplied by $I/c$. With those factors, we conclude that both
results agree with each other.

Using Eq. (\ref{F_rho_inside}) and (\ref{force_in_circular_beam}) we
can also find the relative difference between the force and its
average value
    \be
    {F_\rho-\bar{ F}_\rho \over  \bar{ F}_\rho }=
    \frac{1-2r^2 / a^2 }{4\ln{(8R /a)}-3}.
    \ee
For a thin bunch, when $\ln(8R/a) \gg 1$, the variation of the force
in the cross section is relatively small.

At this point, it is instructive to consider the transverse particle
motion in a coasting beam under the influence of the centrifugal
force. Such motion in the horizontal plane, $z=0$, is governed by the
following equation
    \be
    x''+K x = {e F_\rho (x) \over
    E}  + {1\over R(s)}{\Delta E \over E},
    \ee
where $K$ is the external focusing, and $\Delta E$ is the particle
energy variation arising due to the potential inside the beam,
$\Delta E = -e\phi(x)$,
    \be
    x''+K x = {e  \over
    E}\left(F_\rho  - {\phi\over R}\right).
    \ee
For $z=0$, from Eqs. (\ref{phi0}) and (\ref{F_rho_inside}) we have
    \bea
    \phi&=&\mu\left(1- \frac{x^2}{ a^2}\right)+2\mu\ln{8R \over a},\\
    \nonumber
    F_\rho&=&
    {\mu \over R }\left(2\ln{8R \over a}-1-{x^2\over a^2}\right).
    \eea
We see that the difference
    \begin{align}\label{eq:111}
    F_\rho
    -
    \frac{\phi}{R}
    =
    -
    2\frac{\mu}{R}
    \end{align}
and does not depend on $x$.

%
\section{Short bunch}
%

In this section we will calculate the effective force $\tilde F_\rho \equiv F_\rho-\phi/R$ for a short bunch using a combination of the Lienard-Wiechert fields for a point charge and the result found in the previous section for a coasting beam. We will assume that the bunch density is constant in the cross section within the radius $\sigma_r=a$, and the longitudinal charge distribution per unit length is given by $\mu(s)$ with the rms bunch length $\sigma_z$.

To find the effective force  $\tilde F_\rho$ acting on unit charge in the bunch at point $s=s_0$, we select a small slice of the bunch of length $\Delta s$, such that $\sigma_z \gg \Delta s \gg a$, with a local density $\mu(s_0)$, and calculate the contribution to the force separately from the slice and from the rest of the beam \footnote{In this derivation we assume that the length of the bunch is much larger than its transverse size, $\sigma_z \gg a$. This, however, is not a necessary requirement, and our result is also valid in the limit $\sigma_z \lesssim a$.}. First, let us find the contribution to the force from the bunch excluding the slice. This can be done by integrating the contribution from the part of the bunch outside of the region $[s_0 - \Delta s/2,s_0 + \Delta s/2]$ and treating the beam in this region as infinitely thin, $\sigma_r \to 0$. As was pointed out at the end of the second section, in this approximation, $\tilde F_\rho$ is equal to zero, and hence this contribution vanishes.

To find the contribution to the force from the slice itself, we will use the following approach. Consider  a circular dc beam of constant linear density $\mu=\mu(s_0)$ and  select a slice $\Delta s$ with its center located at $\psi=0$. From the previous section, we know that the effective force in this case does not depend on position and is given by Eq.~(\ref{eq:111}). If we subtract from this force the contribution $\tilde F_\rho$ of the part of the circle external to the slice, that is the part occupying the region $\Delta \psi/2 <\psi<2\pi-\Delta \psi/2$, where $\Delta \psi = \Delta s/R$, we will find the force of the slice itself. However, as was explained in the previous paragraph, the contribution from the slices of the dc beam outside of the range $-\Delta \psi/2 <\psi<\Delta \psi/2$ is zero. So we come to the conclusion that for a short bunch (with the bunch length $\sigma_z\ll R$) the effective force is given by the local value of the beam linear density,
    \begin{align}\label{eq:36}
    \tilde F_\rho(s)
    =
    F_\rho
    -
    \frac{\phi}{R}
    =
    -
    2\frac{\mu(s)}{R}
    .
    \end{align}
This result is in agreement with Ref.~\cite{derbenev96s}.

We can formally apply Eq.~\eqref{eq:36} to a point charge $e$, for which $\mu(s)$ is equal to $e\delta(s)$ which gives for the effective force
    \begin{align}
    \tilde F_\rho(s)
    =
    F_\rho
    -
    \frac{\phi}{R}
    =
    -
    2\frac{e}{R}
    \delta(s)
    .
    \end{align}
This equation corrects Eq.~\eqref{eq:8} which is only valid for $s\ne 0$. Eq.~\eqref{eq:36} can be derived from this equation as a convolution of the point-charge force with the bunch distribution $\mu(s)$.

\section{Projected emittance growth}

In this section, we estimate the effect of the transverse force on
the projected emittance growth of a bunch passing through a bend and compare it
with the emittance growth due to the longitudinal CSR force. We start
from the longitudinal CSR wake. When the beam passes through the
magnet, the energy within the bunch changes, and due to the variation
of the energy, the deflection angle $\Delta x'$ for different slices
of the bunch also varies. This variation is given  by the following
formula,
    \be
    \Delta x'(s)={\theta w_{\rm
    bunch}(s)Ne^2L_b \over 2E},
    \ee
where $\theta$ is the deflection angle for the nominal energy,
$\theta \approx L_b/R$, $L_b$ is the length of the bend, and $E$ is the beam energy. The variance of the deflection angle is proportional to the increase of the projected emittance
$\Delta \epsilon_N$ (we assume that $\Delta \epsilon_N \ll \epsilon_N$),
    \be \label{emit_growth}
    \Delta \epsilon_N = {1\over 2}\gamma \beta \langle
    \left(\Delta x'-\langle \Delta x' \rangle \right)^2 \rangle,
    \ee
where $\beta$ is the beta function at the location of the bend \footnote{This formula is valid when the length of the magnet is much shorter than the beta function, $L_b\ll\beta$, which we assume here.}. For a
Gaussian bunch, using Eq. (\ref{bunch_wake}), we find
    \be \label{emit_growth_long_wake}
    \Delta \epsilon_N
    =
    7.5\times10^{-3} \,{\beta \over \gamma}
    \left({Nr_e L_b^2 \over  R^{5/3} \sigma_z^{4/3}} \right)^2.
    \ee

To estimate the emittance  growth due to the {\em transverse} wake,
we note that the transverse force in the bend deflects the slice by
    \be
    \Delta x'(s)
    =
    \frac{1}{E}
    e\tilde{F}_\rho(s)
    L_b
    \ee
where $\tilde{ F}_\rho$ is given by Eq.~\eqref{eq:36} (we recall that $\tilde{F}_\rho$ is the force per unit charge). Again, using Eq. (\ref{emit_growth}) we find
    \be\label{eq:42}
    \Delta \epsilon_N \approx 2.5\times10^{-2}\, {\beta \over \gamma}
    \left({ Nr_e L_b \over  R \sigma_z} \right)^2.
    \ee
The ratio of the emittance growth given by this equation to the emittance growth given by Eq.~\eqref{eq:36} is
    \begin{align}\label{eq:43}
    3.3
    \left(\frac{ R^{2/3}\sigma_z^{1/3} }{L_b  } \right)^2.
    \end{align}
If we introduce the formation length $L_f$ for the CSR wake with the frequencies of the order $c/\sigma_z$,
    $
    L_f
    \approx
    (24\sigma_z R^2)^{1/3}
    $,
then Eq.~\eqref{eq:43} can be written as
    \begin{align}\label{eq:45}
    0.3
    \left(
    \frac{L_f}{L_b}
    \right)^2.
    \end{align}
The steady-state wake that we used in this paper is only valid for magnets that are longer than the formation length, $L_b\gg L_f$. Hence, the emittance growth due to the transverse wake, Eq.~\eqref{eq:42} is automatically much smaller than the emittance growth~\eqref{emit_growth_long_wake} due to the longitudinal wake.

\section{ Conclusion}

We confirmed the result of Ref.~\cite{derbenev96s} that the effective transverse force for a short relativistic bunch traveling on a circular orbit  is given by the following equation
    \be
    \tilde { F}_\rho(s)
    =
    -
    2{\mu(s) \over R}.
    \ee
Although this centripetal force does not contribute to the tune shift in a circular accelerator, it does effect the transverse motion of the beam. As an example, in this paper we considered the emittance growth of a short bunch in a magnetic compressor. In
this case, the ratio of the emittance  growth due to the transverse force and that due to unshielded CSR wake is given by Eq.~\eqref{eq:45}. It is always small in the regime when the steady state CSR wake can be used.

\section{ Acknowledgements}
The author would like to thank Z. Huang and Y. Cai for useful discussions.

This work was  supported by Department of Energy contract DE-AC03-
76SF00515.


\bibliography{\string~/gsfiles/Bibliography/master}

\end{document}